\documentclass[aps,twocolumn,prd,preprintnumbers, superscriptaddress, 10pt]{revtex4-1}
\usepackage{amsmath,amsfonts,amssymb}
\usepackage{bm}
\usepackage[utf8]{inputenc}
\usepackage{lmodern}
\usepackage{mathrsfs}
\usepackage{mathtools}
\usepackage{tensor}
\usepackage{amsthm}
\usepackage{tikz-cd}

\newcommand{\be}{\begin{equation}}
\newcommand{\ee}{\end{equation}}

\newcommand{\ba}{\begin{aligned}}
\newcommand{\ea}{\end{aligned}}

\newcommand{\cA}{\mathcal{A}}
\newcommand{\cB}{\mathcal{B}}

\newcommand{\cG}{\mathcal{G}}

\newcommand{\cK}{\mathcal{K}}

\newcommand{\cN}{\mathcal{N}}
\newcommand{\cO}{\mathcal{O}}

\allowdisplaybreaks

\usepackage[hidelinks]{hyperref}

\usepackage{comment}
\usepackage[draft,inline,nomargin,index]{fixme}

\FXRegisterAuthor{kr}{akr}{\color{orange}KR}

\begin{document}

\title{\boldmath Non-Abelian symmetry operators from hanging branes in $AdS_5 \times S^5$}
\author{Ibrahima Bah}\affiliation{William H.~Miller III Department of Physics and Astronomy, Johns Hopkins University, 3400 North Charles Street, Baltimore, Maryland 21218, USA}
\author{Federico Bonetti}
\affiliation{Departamento de Electromagnetismo y Electr\'onica, Universidad de Murcia, Campus de Espinardo, 30100 Murcia, Spain}
\author{Mufaro Chitoto}
\affiliation{William H.~Miller III Department of Physics and Astronomy, Johns Hopkins University, 3400 North Charles Street, Baltimore, Maryland 21218, USA}
\author{Enoch Leung}
\affiliation{Max Planck Institute for Mathematics in the Sciences, Inselstraße 22, 04103 Leipzig, Germany}


\begin{abstract}
    We investigate the holographic realization of topological operators for continuous non-Abelian symmetries in quantum field theories. As a concrete case study, we focus on type IIB string theory on $AdS_5 \times S^5$ which admits an $SO(6)$ isometry, dual to the R-symmetry in 4D $\mathcal{N}=4$ super-Yang-Mills theory. We argue that symmetry operators for continuous symmetries are generally realized by bound states of D5-branes and Kaluza-Klein (KK) monopoles hanging from the conformal boundary. Together they account for the contributions to Gauss's law constraints from the $G_5$ flux and the Einstein-Hilbert term respectively. We also demonstrate how the D5-KK bound state measures the representation of the end point of Wilson lines constructed by D3-branes.
\end{abstract}

\maketitle
%
%
%
%
%
%
\section{Introduction}
The study of symmetries in quantum systems has been revolutionized by the development of generalized symmetries, where symmetry data is encoded in the topological sector of quantum field theories~\cite{Gaiotto:2014kfa}. Holography and string theory have been a driving force, providing explicit realizations of topological operators via bulk degrees of freedom~\cite{Apruzzi:2022rei,GarciaEtxebarria:2022vzq,Heckman:2022muc,Heckman:2022xgu,Etheredge:2023ler,Dierigl:2023jdp,Bah:2023ymy,Apruzzi:2023uma,Cvetic:2023pgm,Baume:2023kkf,Yu:2023nyn,DelZotto:2024tae,Hu:2024zvz,Argurio:2024oym,Zhang:2024oas,Braeger:2024jcj,Franco:2024mxa,Gutperle:2024vyp,Knighton:2024noc,Fernandez-Melgarejo:2024ffg,Bergman:2024its,Bonetti:2024etn,Christensen:2024fiq,Caldararu:2025eoj}, and partially inspiring symmetry topological field theories (SymTFTs)~\cite{Witten:1998wy,Bah:2019jts,Bah:2019rgq,Ji:2019jhk,Bah:2020jas,Bergman:2020ifi,Bah:2020uev,Gaiotto:2020iye,Bah:2021brs,Apruzzi:2021nmk,Bergman:2022otk,Freed:2022qnc}, a powerful framework that defines
the action of topological symmetries, and generalizes the action of groups by their representations. Together with fusion categories~\cite{Bhardwaj:2017xup,Thorngren:2019iar,Thorngren:2021yso,Bhardwaj:2022yxj,Muller:2025ext,Bhardwaj:2024xcx,Bah:2025oxi,Kim:2025zdy}, this has yielded deep insight for finite symmetries, both invertible and noninvertible, elucidating fundamental aspects of quantum systems across condensed matter, quantum information, and high-energy physics. For continuous symmetries, however, especially non-Abelian ones, an analogous understanding is still a central unresolved problem~\cite{Garcia-Valdecasas:2023mis,Cvetic:2023plv,Bergman:2024aly,Waddleton:2024iiv,Cvetic:2025kdn,Calvo:2025usj,Najjar:2025htp,Calvo:2025kjh} (see also~\cite{C_rdova_2022}): the categorical structure is lacking, and a SymTFT construction remains an open challenge~\cite{Brennan:2024fgj,Antinucci:2024zjp,Bonetti:2024cjk,Antinucci:2024bcm,Arbalestrier:2025poq,Bonetti:2025dvm,Jia:2025jmn}.

It is natural to expect that the holographic perspective will illuminate the topological nature and action of continuous symmetries. Indeed, a SymTFT is fundamentally a bulk-boundary system, precisely the structure that holography is naturally built to describe. A concrete incarnation of the above challenge in holography is how to measure charges of boundary operators using bulk physics, i.e.~how symmetry operators constructed from bulk degrees of freedom act on boundary operators. Answering this question is important for two reasons: 1) it would provide direct input toward a SymTFT for continuous symmetries, and 2) it is independently a fundamental open problem in AdS/CFT, probing the interplay between gauge symmetry, topology, and brane dynamics.

In this paper, we solve this problem in the prime example of holography: type~IIB string theory on $AdS_5 \times S^5$, dual to 4D $\mathcal{N}=4$ super-Yang-Mills (SYM). 
From the low-energy effective action \cite{Cvetic:2000dm,Pilch:2000ue,Cassani:2010uw,Liu:2010sa,Gauntlett:2010vu,Baguet:2015sma,Gunaydin:1984qu,Pernici:1985ju,Gunaydin:1985cu}, we construct the symmetry operators via Gauss's law constraints and realize them explicitly in string theory as  bound states of Bogomol'nyi--Prasad--Sommerfield (BPS) D5-branes and KK monopoles anchored to the boundary in a U-shaped configuration, whose width provides the necessary regulator for continuous symmetry operators \cite{Bah:2024ucp}.
We characterize the continuous parameters labeling symmetry operators, determine their non-Abelian fusion rules, and demonstrate charge measurement via Hanany-Witten transitions.

While our construction is carried out in a specific setup, the methods extend naturally to general holographic settings~\cite{Bah:2026gcf}. In particular, the explicit identification of the bulk degrees of freedom that realize symmetry operators and their fusion provides a setup to formulate a SymTFT for continuous non-Abelian symmetries that we explore elsewhere.

\section{Low-energy Description}
\subsection{Setup}
The holographic dual of 4D $\cN = 4$ $SU(N)$ SYM is type IIB string theory on $AdS_5 \times S^5$ with $N$ units of $G_5$ flux.
Type IIB supergravity admits a consistent truncation on $S^5$ 
\cite{Cvetic:2000nc,Pilch:2000ue,Cassani:2010uw,Liu:2010sa,Gauntlett:2010vu,Baguet:2015sma}
to 5D maximal  $SO(6)$ gauged supergravity  \cite{Gunaydin:1984qu,Pernici:1985ju,Gunaydin:1985cu}. 
For our purposes
we do not need the full uplift formulas, and focus on the $SO(6)$ gauge fields, not considering fluctuations of the scalar fields or the external metric. 

For the effective action for the $SO(6)$ gauge fields,
we follow the presentation in \cite{Cvetic:2000nc}.
The effective action is
\be \label{eq_eff_action}
S_{\rm eff} = \int_{AdS_5} \bigg[
- \frac {1}{4 g^2} F_{ab} \wedge * F^{ab}
- k {\rm CS}_5
\bigg]  \ . 
\ee 
The $SO(6)$ field strength is 
$F^{ab} = dA^{ab} + A^{ac} \wedge A_c{}^d$, where 
$A^{ab} = - A^{ba}$ and $a$, $b$, etc.~are vector indices of $SO(6)$.
The Chern-Simons 5-form satisfies $d{\rm CS}_5 = \tfrac{1}{384 \pi^2} \epsilon_{abcdef} F^{ab} \wedge F^{cd} \wedge F^{ef}$.
The 
gauge coupling
is $\frac{1}{g^2} = \frac{N^2}{8\pi^2 L}$
and the Chern-Simons level is 
$k=N^2-1$~\cite{Freedman:1998tz}.
Here  $L^4 = 4\pi g_s N (\alpha')^2$.
While  $g^{-2}$ receives 
$1/N$ corrections,
the value of $k$ is exact, as it captures the R-symmetry 't Hooft anomaly of   4D $\mathcal N = 4$ SYM   \cite{Witten:1998qj}.

The gauge coupling constant $g^{-2}$
originates from two terms in the type IIB pseudoaction:  the Einstein-Hilbert (EH) term, 
and the kinetic term for the flux $G_5$. The two contributions are equal up to a factor 2 \cite{Barnes:2005bw},
\be 
\frac{1}{g^2} = \frac{1}{g^2_{\rm EH}} + \frac{1}{g^2_{\rm flux}}
 \ , \quad 
\frac{1}{g^2_{\rm flux}} = \frac{2}{g^2_{\rm EH}}  
\ .\label{eq:effective_gauge_coupling}
\ee

\subsection{Gauss's law constraints} 
Building on \cite{Witten:1998wy,Belov:2004ht}, we analyze the effective action in a Hamiltonian framework where
the radial direction of $AdS_5$ is identified with Euclidean time. The variation of the action with respect to the temporal components of
$A^{ab}$ gives the Gauss's law constraints $\mathbf G_{4ab}=0$, where
\be 
\mathbf G_{4ab} = \tfrac{1}{2g^2} D*F_{ab}
+\tfrac{1}{128 \pi^2} k \epsilon_{ab c_1 c_2 c_3 c_4} F^{c_1 c_2} \wedge F^{c_3 c_4} \ . 
\ee 
Here $D$ is the $SO(6)$ covariant derivative and $*$ is the Hodge star in $AdS_5$. The 4-form $\mathbf G_{4ab}$ is evaluated on a slice at constant $AdS_5$ radius
\footnote{A purely classical analysis of Gauss's law constraints suffices for our purposes. The quantum treatment, in the presence of topological terms in the action, 
requires care~\cite{Belov:2004ht}. }.

Gauss's law constraints generate bulk gauge transformations. A bulk gauge transformation that is nontrivial near the conformal boundary of $AdS_5$ implements a global symmetry.
Thus, we can read off the symmetry generators by taking Gauss's law constraints and evaluating them on a slice at constant $AdS_5$ radius, taking the limit in which we approach the conformal boundary. 
In the standard AdS/CFT dictionary, the gauge fields have Dirichlet boundary conditions $A^{ab}=0$.  Hence,
in this boundary limit, Gauss's law constraints simplify to $\mathbf G_{4ab} = \tfrac{1}{2g^2} d*F_{ab}$.
We may now consider a linear combination 
$\theta^{ab} \mathbf G_{4ab}$ with constant  $\theta^{ab}$ coefficients, and integrate it on a boundary 4-chain $\cB^4$ with
$\partial \cB^4 = M^3$. Applying Stokes's theorem, we get the expected symmetry operator from the low-energy analysis:
\be 
U(\theta;M^3) = \exp \bigg(  \frac{i}{2g^2} \, \theta^{ab} \int_{M^3}  *F_{ab} 
\bigg) \ . \label{eq:symmetry_operator}
\ee 

In \eqref{eq:symmetry_operator}, $\ast F$ 
and $\theta$ take values in $\mathfrak{so}(6)$ and the dual $\mathfrak{so}(6)^*$ respectively. 
Since the Cartan-Killing metric gives an isomorphism $\mathfrak{so}(6)^* \cong \mathfrak{so}(6)$, we will henceforth implicitly identify $\mathfrak{so}(6)^*$ and $\mathfrak{so}(6)$.
$U(\theta;M^3)$ implements the
$SO(6)$ global symmetry transformation associated to the group element $\exp \theta$, 
where $\exp: \mathfrak{so}(6) \to SO(6)$ is the  exponential map,
which is surjective.

\subsection{Wilson line operators}
The action \eqref{eq_eff_action} admits standard Wilson loop operators ${\rm Tr}_{\mathbf R} {\rm Pexp} \int (-A)$ 
labeled by 
a unitary irreducible representation
(irrep) $\mathbf R$ of $SO(6)$.
Let us recall an alternative presentation of
a Wilson loop, based on a quantum mechanical model living on the support $M^1$ of the Wilson loop and coupled to the 5D gauge field $A^{ab}$.
This is based on 
coadjoint orbits and their quantization \cite{Balachandran:1977ub,Alekseev:1988vx,Diakonov:1989fc,Stone:1988fu,Alvarez:1989zv},
see also \cite[Part 4, Sec.~VII.7]{deligne1999quantum}, \cite{Beasley:2009mb,Tong:2014yla}.
The 1D action reads
\be \label{eq_1d_action}
S_{\rm WL} =  \int_{M^1} \tfrac 12
q_{ab} ( \chi^{-1} A \chi + \chi^{-1} d\chi )^{ab} \ ,
\ee 
with 
constants $q_{ab}=q_{[ab]}$,
$\chi^a{}_b$ an $SO(6)$-valued scalar field localized on $M^1$ and $(\chi^{-1} d\chi )^{ab}=(\chi^{-1})^a{}_c d\chi^c{}_d \delta^{db}$.
In \eqref{eq_1d_action}, $A$ is implicitly pulled back from
$AdS_5$.

The only nonzero independent parameters $q_{ab}$ are 
$q_{12}$, $q_{34}$, $q_{56}$ and are integers.
They encode 
the irrep labeling the Wilson line. This will be useful in discussing charge measurement later on. 
We refer the reader to Appendix \ref{App:WilsonLines} for justifications of the statements of this paragraph, which are standard \cite{Balachandran:1977ub,Alekseev:1988vx,Diakonov:1989fc,Stone:1988fu,Alvarez:1989zv,deligne1999quantum,Beasley:2009mb,Tong:2014yla,kirillov2004lectures}.

Consider the end point $\cO$ of a Wilson line in the irrep $\mathbf R$ ending on the conformal boundary of $AdS_5$. The end point transforms as $\cO \to D_{\mathbf R}(g) \cO$ 
under a global $SO(6)$ transformation with parameter $g$.
The map $D_{\mathbf R}$ is the homomorphism sending an abstract group element in  
$SO(6)$ to the matrix that represents it in the irrep~$\mathbf R$.
An $SO(6)$ transformation for $\mathcal{O}$ 
with $g=\exp \theta$
is implemented by the operator  $U(\theta; M^3)$ of the form \eqref{eq:symmetry_operator}, by moving $U(\theta; M^3)$ past
the Wilson line end point \footnote{
Indeed, recall that:
(i) the operator $U(\theta; M^3)$ is constructed from Gauss's law constraints;
(ii) the latter implement bulk gauge transformations; (iii) the global $SO(6)$ action 
originates 
as 
a special instance of a gauge transformation, whose parameter is nontrivial at the  boundary. }.

\begin{figure}[t!]
    \centering
    \includegraphics[scale=0.2]{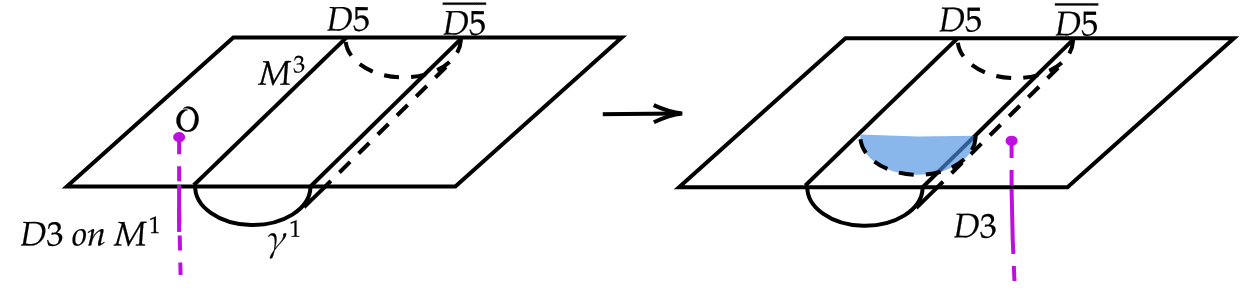}
    \caption{A D5-brane  hangs from the conformal boundary along  $\mathcal{\gamma}^1$ and wraps $M^3$. A D3-brane ends on the conformal boundary on $\mathcal{O}$ and extends along $M^1$ into the $AdS_5$ bulk. 
    A D5-D3 Hanany-Witten move creates an F1-string (blue surface). }
    \label{fig:unlinking}
\end{figure}

\subsection{Wilson lines from D3-branes}
States charged under the $SO(6)$ R-symmetry can be realized using wrapped D3-branes   \cite{Rey:1998ik,McGreevy:2000cw,Grisaru:2000zn,Drukker:2005kx,Benvenuti:2006xg,Gomis:2006sb,Gomis:2006im}. We now revisit this construction to show how the effective action for a D3-brane, with three directions compactified on $S^5$, can reproduce the 1D action
\eqref{eq_1d_action} for a Wilson line in $AdS_5$.

Let $x^\mu$ and the constrained $y^a$ denote coordinates on $AdS_5$ and $S^5$ respectively. The D3-brane world volume action contains scalar fields
$X^\mu(\tau,\sigma^i)$,
$Y^a(\tau, \sigma^i)$, where $\tau$, $\sigma^i$
($i=1,2,3$) are world volume coordinates.
We consider a profile for the scalar fields
$X^\mu$, $Y^a$ of the form $X^\mu(\tau,\sigma^i) = X^\mu(\tau) \ , \; 
Y^a(\tau, \sigma^i) = \chi^a{}_b(\tau) \, Y_{(0)}^b(\sigma^i)$.
The functions $X^\mu(\tau)$ parametrize the D3-brane external support $M^1 \subset AdS_5$ and $Y_{(0)}^a(\sigma^i)$ 
describe the D3-brane internal profile.
The latter wraps $w$ times
around the 
maximal $S^3 \subset S^5$ defined by $y^5=0=y^6$.
We include fluctuations of the internal profile of the brane, associated to
an  $SO(6)$ rotation depending on external spacetime, encoded
by the $\tau$-dependent 
$SO(6)$ matrix~$\chi^a{}_b(\tau)$.

To study the topological couplings on the D3-brane, particularly the $C_4$ term in the Wess-Zumino action,
we pull back the bulk $C_4$ to the D3-brane world volume using the embedding scalars $X^\mu(\tau,\sigma^{i}), Y^a(\tau,\sigma^i)$.
We report the derivation in
\cite{Bah:2026gcf}.
The result is 
the action \eqref{eq_1d_action} 
where the only nonzero  $q_{ab}$ components are $q_{56} =Nw =-q_{65}$.

In general, for any pair of indices $(a,b)$ with $a<b$,
we define $(\Sigma^{3})^{ab}$
as the $S^3 \subset S^5$ determined by $y^a = 0 = y^b$. We also set
$(\Sigma^{3})^{ab} = -(\Sigma^{3})^{ba}$ for $a>b$ and $(\Sigma^{3})^{aa}=0$. With this notation,
the D3-brane wraps 
$M^1 \times \Sigma^3$, where $\Sigma^3 = \tfrac 12 n_{ab} (\Sigma^3)^{ab}$ with arbitrary integer wrapping coefficients~$n_{ab} = -n_{ba}$.
The result is still given by
\eqref{eq_1d_action} with
$q_{ab} = N n_{ab}$. Which components $n_{ab}$ are nonzero translates to different directions in the Lie
algebra $\mathfrak{so}(6) $. The overall factor of $N$
is familiar from the study of
maximal giant graviton states realized by D3-branes
\cite{McGreevy:2000cw, Grisaru:2000zn,   Benvenuti:2006xg}.

In \cite{Bah:2026gcf} we show that the D3-brane effective action Dirac-Born-Infeld (DBI) term is subleading and does not contribute to \eqref{eq_1d_action}.

\section{Symmetry Operators from Hanging Brane Bound States}
As described around \eqref{eq:effective_gauge_coupling}, there are two contributions to Gauss's law constraints $\mathbf{G}_{4ab}$ from the type IIB pseudoaction. We argue that the appropriate stringy configuration which reproduces the full symmetry operator \eqref{eq:symmetry_operator} is a bound state of a D5-brane and a KK monopole corresponding, respectively, to the self-dual $G_5$ flux and the EH action contributions. For both we can identify the 
physical origin of 
 the $\mathfrak{so}(6)$-valued parameter $\theta$ in \eqref{eq:symmetry_operator}. 

We take the world volume
of the D5-KK bound state to be 
$\gamma^1  \times \Sigma^2 \times M^3$. Here $\gamma^1$ is an arc in $AdS_5$ whose end points lie on the conformal boundary at $z=0$ ($z$ denotes the radial coordinate), 
$M^3$ is a 3D subspace of the conformal boundary, and
$\Sigma^2$ is a 2D subspace of $S^5$.
The hanging brane configuration is motivated by \cite{Bergman:2024aly,Calvo:2025kjh} and 
does not depend on the choice of a metric on $M^3$ \cite{Bah:2026gcf}, so in this sense the operator is topological.

To define $\Sigma^2$, we pair the six real constrained coordinates $y^a$ into three complex coordinates, and express $S^5$ as a Hopf fibration over $\mathbb {CP}^2$. Then, 
$\Sigma^2$ can be thought of as a copy of $\mathbb{CP}^1$
inside $\mathbb{CP}^2$.
The choice of complex coordinates can be parametrized by integers $m_{ab} = -m_{ba}$. 
Different choices correspond to different $U(1)$
subgroups of $SO(6)$. Taking products of elements from these subgroups we can construct any element in $SO(6)$;
this is akin to the standard Euler angle parametrization of $SO(3)$ rotations.
See Appendixes \ref{App:SymOp} and \ref{App:Internal} for details. 

\subsection{Hanging D5-brane}
Let us study the effective action
for a D5-brane on $\gamma^1 \times \Sigma^2 \times M^3$, focusing on  topological couplings. The 7-form anomaly polynomial arising from the Wess-Zumino action of the D5-brane contains  the term $2\pi i \int_{N^7} f_2 \wedge G_5$ on an auxiliary 7-manifold $N^7$ whose boundary is the D5-brane world volume. $f_2=da_1$ is the field strength of the Chan-Paton gauge field on the D5-brane,
normalized to have integral periods.
Since $G_5$ is closed on shell, this term can be rewritten as $2\pi i \int_{\gamma^1 \times \Sigma^2 \times M^3} a_1 \wedge G_5$, using Stokes's theorem (with corners). We may then consider a flat Chan-Paton connection with a nontrivial  holonomy along~$\gamma^1$, 
\begin{equation}
    \alpha_\text{D5} = 2\pi \int_{\gamma^1} a_1 \, .\label{eq:D5_holonomy}
\end{equation}
Here $\alpha_\text{D5}$ is an angular variable with period $2\pi$.
This can be seen from
the $a_1 \wedge G_5$ coupling
and the fact that $G_5$ has integral periods, or alternatively from the
fact that $\alpha_{
D5}$ is the parallel transport of a   $U(1)$ connection.
We impose Dirichlet boundary conditions on $a_1$ at the end points of $\gamma^1$ to ensure gauge invariance.
Next we evaluate $\int_{\Sigma^2 \times M^3} G_5$ using \eqref{eq_G5_integral} to get
\begin{equation}
    U_\text{D5}(\alpha_\text{D5},m^{ab};M^3) = \exp\bigg(\frac{i \, \alpha_\text{D5} \, m^{ab}}{2g_\text{flux}^2} \,  \int_{M^3} \ast F_{ab}\bigg) \ .
\end{equation}
\subsection{Hanging KK monopole}
We now turn to the KK monopole, whose effective action was derived in \cite{Eyras:1998hn} (see references therein). The world volume bosonic fields include four real scalars associated with the  four transverse directions
 to the KK monopole, two compact scalars whose 1-form field strengths will be denoted as $\cK_1$ and $\widetilde \cK_1$, and one chiral 2-form.  

As in the D5-brane analysis, one can derive the 7-form anomaly polynomial via descent \cite{Eyras:1998hn,Bah:2026gcf}. The term of interest is $ 2\pi i \int_{N^7} \widetilde \cK_1 \wedge \cK_1 \wedge G_5$,
where $d\cK_1=d\widetilde \cK_1=0$ by the Bianchi identities. We can regard $\widetilde \cK_1 \wedge \cK_1$ as the 2-form field strength of a composite $U(1)$ gauge field $v_1$, such that $dv_1 = \widetilde \cK_1 \wedge \cK_1$, and then write the relevant term in the form $
2\pi i \int_{\gamma^1 \times \Sigma^2 \times M^3} v_1 \wedge G_5 $. Analogously to $\alpha_{D5}$,
we turn on a flat profile for $v_1$
with a nontrivial holonomy along $\gamma^1$,
\begin{equation}
    \alpha_\text{KK} = 2\pi \int_{\gamma^1} v_1 \, ,
\end{equation}
and impose Dirichlet boundary conditions for $v_1$. 
The integral $\int_{\Sigma^2\times M^3}G_5$ is done in the Appendix \ref{App:SymOp}.
Recalling \eqref{eq:effective_gauge_coupling}, the resultant operator is
\begin{equation}
 \!\!\!\!   U_\text{KK}(\alpha_\text{KK},m^{ab};M^3) = \exp\bigg(\frac{i \, \alpha_\text{KK} \, m^{ab}}{g_\text{EH}^2}   \int_{M^3} \ast F_{ab}\bigg) \, .
\end{equation}

To reproduce the full symmetry operator \eqref{eq:symmetry_operator} determined by Gauss's law constraints, the D5-KK bound state should satisfy the relation
$\alpha \equiv \alpha_\text{D5} = 2\alpha_\text{KK}$.
This takes into account the contributions from the $G_5$ flux and the EH term. All in all we get the overall operator
\begin{equation}
    U_\text{D5-KK}(\alpha,m^{ab};M^3) = \exp\bigg(\frac{i \, \alpha \, m^{ab}}{2g^2}   \int_{M^3} \ast F_{ab}\bigg) \, ,\label{eq:D5-KK_bound_state}
\end{equation}
where $g^2$ is the  gauge coupling given by \eqref{eq:effective_gauge_coupling}. 
We notice that \eqref{eq:D5-KK_bound_state} matches with the operator   \eqref{eq:symmetry_operator} from the low-energy analysis, with the identification
$\theta^{ab} = \alpha m^{ab}$. We stress again that both the D5-brane and the KK monopole are BPS objects in type IIB string theory.

\section{Charge Measurement from Branes}
The operator realized by the hanging 5-brane measures the charge of the end point of the Wilson line realized by the D3-brane. This can also be  understood  in terms of a Hanany-Witten transition \cite{Hanany:1996ie}. Only the D5-brane inside the D5-KK bound state participates in the Hanany-Witten move.

For definiteness, we take a 
D5-brane wrapping $\Sigma^2 \subset S^5$, extending along $x^0$, $x^1$, $x^2$, and along an arc $\gamma^1$ in the 2D plane spanned by $x^3$ and the $AdS_5$ radial coordinate $z$. The D3-brane wraps $\Sigma^3 \subset S^5$ and extends along $z$.
Initially the D3-brane sits to the left of the hanging D5-brane in the $x^3$ direction. If we bring the D3-brane to the other side of the D5-brane, an F1-string is created. 
Its   world sheet is a semidisk
in the $x^3$-$z$ plane, bounded by  $\gamma^1$ and the conformal boundary of $AdS_5$ \cite{Calvo:2025kjh} (see Fig.~\ref{fig:unlinking}). Internally,
the F1-string is supported on the intersection of $\Sigma^3$, $\Sigma^2$.
We assume $\Sigma^3$ intersects transversely with $\Sigma^2$,
and denote by $\Sigma^3 \cdot \Sigma^2$ their intersection~number.

The F1-string created by the Hanany-Witten transition measures charge. To see this, we focus on the topological couplings on the F1-string. If an F1-string with world sheet $W^2$ ends on a D$p$-brane, it couples to the Chan-Paton gauge field $a_1$ on the 
D$p$-brane,
$S^{\rm top}_{\rm F1} \supset  2\pi \int_{\partial W^2} a_1$.
The F1-string is suspended between the hanging D5-brane and the $AdS_5$ boundary, therefore, 
$\partial W^2$
consists of the arc $\gamma^1$, and a segment on the  boundary. The coupling of the F1-string to the D5-brane Chan-Paton gauge field happens only along the $\gamma^1$ portion of~$\partial W^2$. This 
yields the exponentiated action $\exp \left[  2\pi i (\Sigma^3 \cdot \Sigma^2) \int_{\gamma^1} a_1 \right]
= \exp\left[  i \alpha (\Sigma^3 \cdot \Sigma^2)\right] $
where we used $\alpha_{D5}=2\pi \int_{\gamma^{1}}a_1$ and 
the effective multiplicity of the F1-string is $\Sigma^3\cdot \Sigma^2 = \tfrac 12 N n_{ab} m^{ab}$, in terms of the integer parameters $n_{ab}$, $m^{ab}$ that specify $\Sigma^3$, $\Sigma^2$, respectively. Identifying 
$q_{ab} = N n_{ab}$
and $\theta^{ab} = \alpha m^{ab}$, from earlier, we can   write $\exp\left[  i \alpha (\Sigma^3 \cdot \Sigma^2)\right]$ in the form
$\exp(\tfrac 12 i \theta^{ab} q_{ab} )$.

We can use Hanany-Witten transitions to probe a given D3-brane with different hanging 5-branes, with various internal profiles.
Each Hanany-Witten move results in a phase factor 
$\exp(\tfrac 12 i \theta^{ab} q_{ab} )$
with the same $q_{ab}$ from the D3-brane, but different $\theta^{ab}$ depending on the internal profile of the 5-brane and  the holonomy of its Chan-Paton gauge field. 
This set of phase factors allows us to  reconstruct the values of the charges $q_{ab}$,
and thereby determine the irrep $\mathbf R$ associated to the D3-brane.

The topological couplings of the F1-string include a coupling to the Neveu-Schwarz--Neveu-Schwarz (NSNS) 2-form $B_2$ that, in our setup, is trivial in the background and is also not sourced by the 5-brane insertion. Thus, this does not contribute.
We also expect that 
the nontopological Nambu-Goto world sheet term does not affect the conclusions of this section.

We note that the D3-branes wrapping the maximal $S^3\subset S^5$ correspond to the maximal giant gravitons of \cite{McGreevy:2000cw, Grisaru:2000zn,   Benvenuti:2006xg}. As such, they do not collapse to a point and are stabilized at finite size by the angular momentum carried on $S^5$, which is encoded in the collective-coordinate rotation $\chi$ entering \eqref{eq_1d_action}. This balances the brane tension against the coupling to the background 5-form flux. More generally, although the wrapped sphere is contractible and hence trivial in ordinary homology, the relevant configurations are nontrivial as equivariant cycles, i.e. invariant submanifolds under the relevant $U(1)\subset SO(6)$ isometry action. It is this equivariant data, rather than the (trivial) homology class, that determines the charge and topological properties of the associated operators; we develop this viewpoint in detail in \cite{Bah:2026gcf}. The same structure applies to the hanging 5-branes realizing the symmetry operators: their internal supports are likewise invariant submanifolds under the same isometry action, so that their topological properties are controlled by this equivariant data rather than by the (trivial) homology class.

\section{Fusion of Hanging Branes}
\label{sec:fusion-section}
An operator $U(\theta; M^3)$ of the form \eqref{eq:symmetry_operator}, realized as a D5-KK bound state \eqref{eq:D5-KK_bound_state}, implements the $SO(6)$ transformation $g=\exp \theta$. We denote it as $U_g$.
The parallel fusion of such operators is group-like,
$U_g U_{g'} = U_{gg'}$. This can be recovered from
\eqref{eq:effective_gauge_coupling}, the Baker-Campbell-Hausdorff (BCH) formula, and the algebra of $\int_{M^3}*F$. To describe the latter we define
$X_{ab} =\frac{i}{2g^2} \int_{M^3} \ast F_{ab}$ and write 
$[X_{ab},X_{cd}] = \tfrac 12f_{ab,cd}{}^{ef} X_{ef}$,
$f_{ab,cd}{}^{ef} = 8 \delta_{[a}{}^{[e} \delta_{b][c} \delta_{d]}{}^{f]}$. See Appendix \ref{App:NonComm} for more details.

Consider $g=\exp(\alpha m)$,
$g'=\exp(\alpha' m)$ for the same $m$ in $\mathfrak{so}(6)$. Fusion amounts to summing the holonomies $\alpha$ and $\alpha'$.
The two D5-KK bound states share
the same internal support
but  have independent pairs of world volume gauge fields $(a_1,v_1)$ and $(a_1',v_1')$ respectively.
We focus  on the Chan-Paton gauge field $a_1$. The argument for the KK monopole gauge field $v_1$ follows analogously.

\begin{figure}[t!]
    \centering
    \includegraphics[scale=0.2]{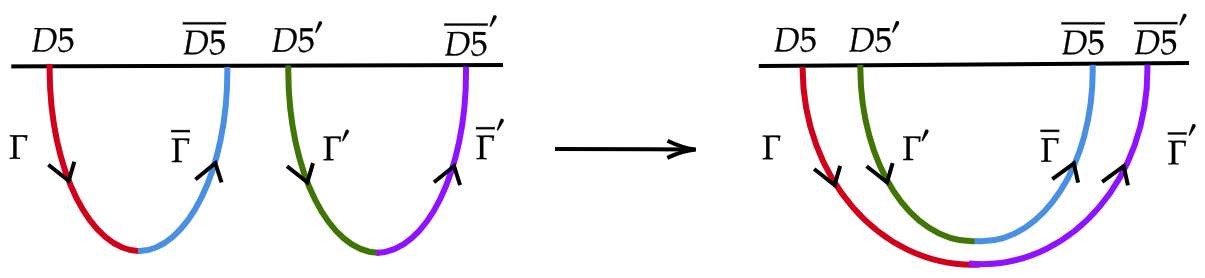}
    \caption{Two hanging branes on $\Gamma \sqcup \overline{\Gamma}$ and $\Gamma^\prime \sqcup \overline{\Gamma}^\prime$ split and recombine into a hanging two-brane-stack on $\Gamma \sqcup \overline{\Gamma}^\prime$ and $\Gamma^\prime \sqcup \overline{\Gamma}$.}
    \label{fusion2}
\end{figure}

Regard $\gamma^1 = \Gamma \sqcup \overline{\Gamma}$ as a disjoint union of two half-arcs with opposite orientations, on which the D5-brane and the $\overline{\text{D}5}$-brane in the hanging brane solution are respectively supported. The holonomy is then the sum $\alpha_\text{D5} = \alpha_\Gamma - \alpha_{\overline{\Gamma}}$ where $\alpha_\Gamma = \int_\Gamma a_1$. As we bring two such hanging branes toward each other, each of them (momentarily) breaks up into a D5-$\overline{\text{D}5}$ pair, then the two outer components recombine to form a single hanging D5-brane on $\Gamma \sqcup \overline{\Gamma}'$ (see Fig.~\ref{fusion2}). It has the same orientation as the original arcs $\gamma^1$ and $(\gamma^1)'$, so the associated holonomy is $\widetilde{\alpha} = \alpha_\Gamma - \alpha'_{\overline{\Gamma}'}$. To ensure the inner components also recombine in a way compatible with their orientations, we see that they must move past each other and then recombine into $\Gamma' \sqcup \overline{\Gamma}$, with holonomy $\widetilde{\alpha}' = \alpha'_{\Gamma'} - \alpha_{\overline{\Gamma}}$.

The result is a stack of two hanging D5-branes whose effective holonomy is $\widetilde{\alpha} + \widetilde{\alpha}' = \alpha + \alpha'$, as desired, which is associated with the center-of-mass mode of the stack of D5-branes \cite{Bah:2023ymy}. Due to the tracelessness of $\mathfrak{su}(2)$-valued gauge fields, the enhanced non-Abelian component of the Chan-Paton gauge field does not contribute to the topological coupling of interest. We can similarly keep fusing more hanging branes as long as they share the same world volume. This operation is associative. See~\cite{Bah:2026gcf} for details on the brane dynamics underlying fusion.

If instead we have $g=\exp(\alpha m)$,
$g'=\exp(\alpha' m')$ with $m$ and $m'$ 
 nonzero along independent factors in the Cartan subalgebra $\mathfrak{u}(1)^{\oplus 3}$, then they generate simultaneous rotations in orthogonal planes in $SO(6)$. In this sense, the two D5-KK bound states are supported on orthogonal 2D subspaces $\Sigma^2$. What appears as a parallel fusion in the low-energy effective theory is really two noncoincident D5-KK bound states in the full 10D string theory background sharing the same external support $\gamma^1 \times M^3$. This is why we can build a symmetry operator labeled by a generic element in the maximal torus, parametrized by three angles $(\alpha,\alpha',\alpha'') \in \mathfrak{u}(1)^{\oplus 3}$. Symmetry operators ``supported'' along orthogonal $\mathfrak{u}(1)$ directions commute, and so the associated holonomies do not directly add.

In the most general scenario, without imposing any
condition on $m$, $m'$,
 the internal supports  of the two D5-KK bound states are neither coincident nor orthogonal. Fusion thus involves 
higher-order terms in the BCH formula, which determine the internal support of the final product when regarded as a single D5-KK bound state.

\section{Discussion and Outlook}
We have constructed the first explicit brane realization of continuous non-Abelian symmetry operators in holography, as bound states of hanging D5-branes and KK monopoles in $AdS_5 \times S^5$. The full $SO(6)$ R-symmetry operator, its continuous parameters, and non-Abelian fusion rules can be derived from the brane dynamics in the UV and matched with low-energy expectations from Gauss's law. Charge measurement is realized via Hanany-Witten transitions. Our construction raises several questions. The symmetry operators here wrap subspaces defined by Killing vectors, in contrast to finite symmetries where branes wrap internal homological cycles; understanding this distinction should clarify the categorical structure of continuous symmetries. It would also be valuable to find a mechanism beyond Gauss's law that fixes the factor of 2 between D5-brane and KK monopole holonomies. In~\cite{Bah:2026gcf}, we generalize to backgrounds with richer topologies and study the dynamics of hanging branes. More broadly, our results provide concrete holographic data, namely, symmetry operators, their parameters, and fusion, that a SymTFT for continuous non-Abelian symmetries must encode. An emerging lesson is that the mathematics of generalized symmetries is implicitly realized by the physics of branes in string theory; our work takes an important step in this direction, and fully uncovering this correspondence is a compelling goal for the future.

\acknowledgments
We would like to thank Minjae Kim, Nitu Kitchloo, and Thomas Waddleton for illuminating discussions.  I.~B. thanks Zohar Komargodski for the wonderful hospitality at the Simons Center for Geometry and Physics, and for the excellent and lively discussions in all things while completing this work. 
I.~B. and M.~C. are supported in part by the Simons Collaboration on Global Categorical Symmetries and also by the NSF Grant No.~PHY-2412361. F.~B. is supported by 
the Program ``Saavedra Fajardo'' 22400/SF/23
from
Fundaci\'on S\'eneca de la Regi\'on de Murcia.
F.~B. also acknowledges support from
Fundaci\'on S\'eneca de la Regi\'on de Murcia FSRM/10.13039/100007801(22581/PI/24), Espa\~na.

\section*{Data availability}
There are no publicly available research data or software supporting this manuscript. Requests for further information or data should be sent to the authors.

\let\oldaddcontentsline\addcontentsline
\renewcommand{\addcontentsline}[3]{}

\let\addcontentsline\oldaddcontentsline


\appendix
\addtocontents{toc}{\protect\setcounter{tocdepth}{1}}


\section{Type IIB String Theory Setup}\label{App:Type2}
We provide additional details concerning the type IIB holographic setup that gives rise to the desired operators. Type IIB string theory on $AdS_5\times S^5$ with $N$ units of $G_5$ flux is holographically dual to 4D $\cN = 4$ $SU(N)$ super-Yang-Mills.
The 10D metric reads
\be \label{eq_10d_metric}
ds^2_{10} = ds^2_{AdS_5} + L^2 Dy^a Dy_a \ ,
\quad 
Dy^a = dy^a + A^{ab} y_b \ .
\ee 
Here $L$ is the radius of $AdS_5$, $A^{ab} = A^{[ab]}$ are the $SO(6)$ gauge fields,
and $a,b=1,\dots, 6$ are vector indices of $SO(6)$.
The constrained coordinates $y^a$
satisfy $y^a y_a = 1$ and parametrize~$S^5$.
The 5-form flux is given as
\be \label{eq_G5}
G_5 = \cG_5 + *_{10} \cG_5 \, , \,\,\,\,\,\,\,\,\,
\mathcal \cG_5 = N \Big[ 
(V_5)^{\rm g} + \tfrac 12 \tfrac{F^{ab}}{2\pi} \wedge (\omega_{ab})^{\rm g}
\Big] \, ,
\ee 
where $*_{10}$ is the Hodge operator of the metric \eqref{eq_10d_metric}.
In our normalization, $G_5$
has integral periods.
The 5-form $V_5$  and 3-forms $\omega_{ab}$ are defined as 
\be \label{eq_omega_def}
\ba 
V_5 &= \tfrac{1}{\pi^3 \cdot 5!} \epsilon_{abcdef} y^a dy^b \wedge dy^c\wedge dy^d\wedge dy^e\wedge dy^f \ , \\ 
\omega_{ab} &= \tfrac{1}{2\pi^2 \cdot 3!}\epsilon_{abcdef}  y^c\wedge dy^d\wedge dy^e\wedge dy^f \ . 
\ea 
\ee 
The superscript ``g'' in \eqref{eq_G5}
stands for the replacement
$dy^a \to Dy^a$ in a differential form. It captures the coupling to the external gauge fields $A^{ab}$. 
The Hodge dual of $\omega_{ab}$
with respect to the round metric on $S^5$~is 
\be \label{eq_omega_and_Killing}
* \omega_{ab} = \tfrac{1}{4\pi^2} dK_{ab} \ , \qquad 
K_{ab} = 2y_{[a} dy_{b]} \ .
\ee 
The vector fields obtained from the 1-forms $K_{ab}$ by raising their tangent indices with the round metric  are the $SO(6)$ Killing vectors on~$S^5$. It is important to note that, 
\be 
 \int_{S^5} \omega_{ab} \wedge * \omega^{cd} = \frac{1}{3\pi} \, \delta_{[a}{}^c \delta_{b]}{}^d=\frac{4\pi L}{N^2 g^2_{\rm flux}} \, \delta_{[a}{}^c \delta_{b]}{}^d \ .\label{eq:omega_orthogonality}
\ee 
where $g^2_{\rm flux}$ is the coupling defined in \eqref{eq:effective_gauge_coupling}.

\section{Wilson Lines as 1D Quantum Mechanical Models}
\label{App:WilsonLines}
The 1D Wilson line action \eqref{eq_1d_action} is
\be \label{eq_1d_action_repeat}
S_{\rm WL} =  \int_{M^1} \tfrac 12
q_{ab} ( \chi^{-1} A \chi + \chi^{-1} d\chi )^{ab} \ ,
\ee 
where 
$q_{ab}=q_{[ab]}$ are constants,
$\chi^a{}_b$ is an $SO(6)$-valued scalar field localized on $M^1$. 
The 
action \eqref{eq_1d_action}
is invariant under 5D $SO(6)$ gauge transformations
\be \label{eq_A_gauge}
A^{a}{}_b \to ( M  A  M^{-1} 
+ M  dM^{-1} )^a{}_b \ ,
\,\,\,
\chi^a{}_b \to M^a{}_c \, \chi^c{}_b \ ,
\ee 
with parameter $M \in SO(6)$. Moreover,
$e^{i S_{\rm WL}}$
is  invariant under the local transformation
\be 
\chi^a{}_b \to \chi^a{}_c \, h^c{}_b \ , 
\ee 
provided that $q_{ab}\in \mathbb Z$. Here
$h^a{}_b$ is a 0-form on $M^1$ valued in the subgroup $S_q$ of $SO(6)$ that stabilizes $q_{ab}$, i.e.~$S_q = \{g \in SO(6)| (g q g^{-1})_{ab}=q_{ab}\}$.

The parameter $q_{ab}$
determines an element in 
$\mathfrak{so}(6)^* \cong \mathfrak{so}(6)$
which pairs with the 
$\mathfrak{so}(6)$-valued 1-form
$(\chi^{-1} A\chi + \chi^{-1}d\chi)^{ab}$.
We can 
interpret \eqref{eq_1d_action} as a topological sigma model into the coadjoint orbit $SO(6)/S_q$ of $q_{ab}$. 
Integrality of 
$q_{ab}$
ensures integrality of the 
canonical symplectic 2-form on the orbit
\cite{kirillov2004lectures}.

The physics of the action \eqref{eq_1d_action}
depends on the parameter $q_{ab}$ only through its orbit
$SO(6)/S_q$. 
This means that we have the freedom to 
 ``gauge-fix''
$q$ to a convenient representative of its orbit. 
In particular, it is always possible to choose $q$ in the standard Cartan subalgebra of $\mathfrak{so}(6)$, consisting of antisymmetric matrices whose only nonzero independent entries are $q_{12}$, $q_{34}$, $q_{56}$.
After this is achieved, there is a ``residual gauge freedom'' by the action of the Weyl group,
which further allows us to select $q$ inside the fundamental Weyl chamber in the Cartan subalgebra.
More formally, 
it is known that 
each coadjoint orbit intersects the fundamental Weyl chamber in the Cartan subalgebra  at one point
(see e.g.~\cite[Chap.~5, Lemma 3]{kirillov2004lectures}). 
Thus,
coadjoint orbits are in 1-to-1 correspondence with points in the Weyl chamber.

\nocite{hall2000elementary}
Let us use the fact  that $q$ must be integral. This selects a discrete subset of points inside the Weyl chamber, namely, 
dominant integral elements.
We recall that, by virtue of the highest-weight theorem,
the following holds: 
(i) every finite-dimensional irrep of $\mathfrak{so}(6)$ has a unique highest-weight, which is a dominant integral; (ii) 
for any dominant integral element $q$
there exists a unique finite-dimensional irrep of $\mathfrak{so}(6)$ with
highest-weight $q$ \footnote{As far as group representations are concerned, dominant integral 
elements are in 1-to-1 correspondence with isomorphism classes of unitary finite-dimensional irreps of the simply connected $SU(4)$. For $SO(6)$, an analogous statement holds, but the notion of 
integral element has to be replaced by that of analytically integral element, see e.g.~\cite{hall2000elementary}.
While all analytically integral elements
are integral, the converse is not true. 
Integral elements that are not analytically integral 
 correspond to irreps of $SU(4)$ which are not
irreps of~$SO(6)$.
}.
Now the crucial point is that 
the quantization of the classical theory with action \eqref{eq_1d_action} and parameter $q$
yields a finite-dimensional Hilbert space which is precisely isomorphic to the
 irrep with highest-weight $q$, 
 as reviewed e.g.~in \cite[Sec.~IV.1]{Beasley:2009mb}.

All in all, knowledge of the parameter $q$ in the action \eqref{eq_1d_action} contains the same information as the irrep labeling the Wilson line.
\section{Internal profile of the 5-branes}
\label{App:Internal}
The 5-brane bound state that we study is supported on the world volume $M^3\times\gamma^1\times \Sigma^2$, where $M^3\times\gamma^1\subset AdS^5$ and $\Sigma^2\subset S^5$. We will now discuss the internal profile
$\Sigma^2$ of this 5-brane bound state. To this end, we realize $S^5$ as a Hopf fibration over $\mathbb{CP}^2$.
This can be achieved 
by introducing
complex coordinates $w_1$, $w_2$, $z$ via
\be \label{eq_def_z_w1_w2}
z w_1 = y_1 + iy_2 \ , \quad 
z w_2 = y_3 + iy_4 \ , \quad 
z = y_5 + iy_6 \ .
\ee 
We write $z = |z|e^{i\psi}$.
The constraint $y^a y_a =1$
fixes $|z|$ in terms of $w_1$, $w_2$.
The angle $\psi$, with period $2\pi$,
parametrizes the Hopf fiber,
while $w_1$, $w_2$ are affine coordinates on the $\mathbb{CP}^2$ base. The $S^5$ line element reads
\be \label{eq_Hopf}
ds^2_{S^5} = ds^2_{\mathbb C 
\mathbb P^2} + (d\psi + \cA)^2 \ ,
\ee 
where $ds^2_{\mathbb C 
\mathbb P^2}$ is the Fubini-Study metric on $\mathbb{CP}^2$ and  the connection 1-form $\cA$ satisfies $d\cA = 2J$, with $J$ the K\"ahler 2-form on   
$\mathbb C \mathbb P^2$.
We also notice that the 1-form
dual to the Killing vector $\partial_\psi$ is
\be \label{eq_Killing_1form}
k_\psi = d\psi + \cA = \tfrac 12 m^{ab} K_{ab} \ . 
\ee 
In the second equality 
we have expanded $k_\psi$ onto the
1-forms
$K_{ab}$ defined in \eqref{eq_omega_and_Killing}.
Without loss of generality, the expansion coefficients $m^{ab}$
can be taken to be antisymmetric.
They are computed to be
\be \label{eq_m_from_psi}
m^{ab} = 
\left\{
\begin{array}{ll}
+1 & \text{if $(a,b)=(1,2)$, $(3,4)$, $(5,6)$} \ , \\
-1 & \text{if $(a,b)=(2,1)$, $(4,3)$, $(6,5)$} \ , \\
0 & \text{otherwise}  \ .
\end{array}
\right.  
\ee 
This  can be verified explicitly
using the expression \eqref{eq_omega_and_Killing} for
$K_{ab}$ in  Hopf coordinates.

Having cast $S^5$ as a Hopf fibration,
 we can now characterize $\Sigma^2$: it is the 
2-cycle in  $\mathbb C \mathbb P^2$  which is Poincar\'e dual to~$N \frac{d\cA}{2\pi}$,
\be \label{eq_Sigma2_def}
\Sigma^2 = {\rm PD}_{\mathbb{CP}^2}[N \tfrac{d\cA}{2\pi}] \ .
\ee 
The factor of $N$ is justified below.
We notice the identity
\be \label{eq_dA_omega_identity}
\tfrac{d\cA}{2\pi} \wedge \tfrac{D\psi}{2\pi} = \tfrac 12 m^{ab} \omega_{ab} \ , 
\ee 
with the same coefficients $m^{ab}$ as in
\eqref{eq_m_from_psi}.
One can verify \eqref{eq_dA_omega_identity}
from the expressions \eqref{eq_omega_def}
for the forms $\omega_{ab}$
evaluated in Hopf coordinates.

The identity \eqref{eq_dA_omega_identity} is useful for evaluating integrals over~$\Sigma^2$,
\be \label{eq_poincare_integral}
\int_{\Sigma^2} \lambda_2 = \int_{S^5} N\tfrac{d\cA}{2\pi} \wedge \tfrac{D\psi}{2\pi} \wedge \lambda_2 = \int_{S^5} \tfrac 12 Nm^{ab} \omega_{ab} \wedge \lambda_2 \ , 
\ee 
where $\lambda_2$ is a 2-form. Based on \eqref{eq_poincare_integral} we can  formally write 
$\Sigma^2 = \tfrac 12 Nm^{ab} \Sigma^2_{ab}$,
where integrals over $\Sigma^2_{ab}$ are defined via
\be 
\int_{\Sigma^2_{ab} } \lambda_2 = \int_{S^5}   \omega_{ab} \wedge \lambda_2 \ . 
\ee 

While $\Sigma^2$ is a nontrivial 2-cycle in $\mathbb{CP}^2$, 
it is trivialized by the Hopf fibration (as it must be the case, since $S^5$ has no nontrivial 2-cycles) \footnote{This can be rephrased as follows.
The 2-form $N \tfrac{d\cA}{2\pi}$
is cohomologically nontrivial on the base $\mathbb{CP}^2$. If we pull it back to the total space of the Hopf fibration, however,
it becomes cohomologically trivial,
since
we can write 
$N \tfrac{d\cA}{2\pi} = \tfrac{N}{2\pi} dk_\psi$,
and $k_\psi$ is a globally defined 1-form in the total space.}.

In our construction,  we insert a probe KK monopole. 
We recall that 
the BPS KK monopole of type IIB string theory is a supersymmetric soliton with a 6D world volume. 
One of its four transverse directions 
plays a distinguished role and must be
an isometry of the 10D metric.
In the present setting, this isometry is identified with $\partial_\psi$.
Correspondingly, the KK monopole supports $\Sigma^2$.
Thanks to the probe KK monopole,
we can also wrap a D5-brane on $\Sigma^2$.

\section{Symmetry Operator from the 5-brane}\label{App:SymOp}
The desired symmetry operator comes from the following term of the D5-brane action, $i\alpha_{D5}\int_{\Sigma^2\times M^3} G_5$. The relevant terms are
$G_5 \supset \frac{1}{2}N L^{-1} \frac{\ast F_{cd}}{2\pi} \wedge \ast \omega^{cd}$,
where the $L^{-1}$ factor comes from the $L^2$ in front of the $S^5$ metric in the 10D line element \eqref{eq_10d_metric} when computing $\ast_{10} \cG_5$.
The integral of $\ast \omega^{cd}$
over $\Sigma^2$ is conveniently performed 
using \eqref{eq_poincare_integral} and \eqref{eq:omega_orthogonality}.
The result reads
\begin{equation}
\label{eq_G5_integral}
    \int_{\Sigma^2 \times M^3} G_5 = \frac{1}{2g_\text{flux}^2} \, m^{ab} \int_{M^3} \ast F_{ab} \, ,
\end{equation}
with the same $m^{ab}$ coefficients
as in \eqref{eq_Killing_1form}, satisfying 
\eqref{eq_m_from_psi}.
The factor of $N$ in the Poincar\'e dual \eqref{eq_Sigma2_def}
of $\Sigma^2$ ensures that the right-hand side
has no explicit $N$ factors.
Finally, 
\eqref{eq_G5_integral} combines with $\alpha_\text{D5} = 2\pi \int_{\gamma^1} a_1$, to produce the operator
\begin{equation}
    U_\text{D5}(\alpha_\text{D5},m^{ab};M^3) = \exp\bigg(\frac{i \, \alpha_\text{D5} \, m^{ab}}{2g_\text{flux}^2} \,  \int_{M^3} \ast F_{ab}\bigg)
\end{equation}
from the hanging D5-brane. The computation for the KK monopole follows in the same way to give 
\begin{equation}
    U_\text{KK}(\alpha_\text{KK},m^{ab};M^3) = \exp\bigg(\frac{i \, \alpha_\text{KK} \, m^{ab}}{g_\text{EH}^2} \,  \int_{M^3} \ast F_{ab}\bigg)
\end{equation}
where $g^2_{\text{EH}}$ is the coupling defined in \eqref{eq:effective_gauge_coupling}. These two contributions then combine to give the full symmetry operator 
\begin{equation}
    U_\text{D5-KK}(\alpha,m^{ab};M^3) = \exp\bigg(\frac{i \, \alpha \, m^{ab}}{2g^2}   \int_{M^3} \ast F_{ab}\bigg) \, .\label{boundstateEq}
\end{equation}
\subsection{Generalization of the symmetry operator} Our definition of Hopf coordinates in the Appendix \ref{App:Internal} is subordinate to the choice \eqref{eq_def_z_w1_w2}
 of complex coordinates. The entire construction, however, can be repeated verbatim
for different choices of complex coordinates. Indeed, we can consider
a generalization of \eqref{eq_def_z_w1_w2} of the form
\be \label{eq_new_complex}
\ba
z w_1 &= y_{\sigma_1} + iy_{\sigma_2} \ , & 
z w_2 &= y_{\sigma_3} + iy_{\sigma_4} \ , \\ 
z &= y_{\sigma_5} + iy_{\sigma_6} \ ,
\ea
\ee 
where $\sigma:\{1,\dots,6\} \to \{1,\dots,6\}$ is a permutation and $\sigma_i = \sigma(i)$.
The relations 
\eqref{eq_Killing_1form} and
\eqref{eq_dA_omega_identity},
as well as the final result \eqref{boundstateEq},
still hold, 
albeit with different $m^{ab}$ coefficients. More precisely, 
\eqref{eq_Killing_1form},
\eqref{eq_dA_omega_identity},
and \eqref{boundstateEq}
hold with $m^{ab} = m^{ab}_{(\sigma)}$ where we have 
 introduced the notation
\be \label{eq_m_from_psi_more}
m_{(\sigma)}^{ab} \!=\! 
\left\{
\begin{array}{ll}
\!\!+1 & \text{\small if $(a,b)=(\sigma_1,\sigma_2)$, $(\sigma_3,\sigma_4)$, $(\sigma_5,\sigma_6)$} \ , \\
\!\!-1 & \text{\small if $(a,b)=(\sigma_2, \sigma_1)$, $(\sigma_4,\sigma_3)$, $(\sigma_6,\sigma_5)$} \ , \\
\!\!0 & \text{\small otherwise}  \ .
\end{array}
\right.  
\ee 
For the trivial permutation $\sigma = (123456)$,
$m_{(\sigma)}^{ab}$ equals $m^{ab}$ in \eqref{eq_m_from_psi}.

All in all, repeating our construction with different complex coordinates, 
we have a 5-brane realization of any operator $U(\theta)$,
\be 
U(\theta;M^3) = \exp \bigg(  \frac{i}{2g^2} \, \theta^{ab} \int_{M^3}  *F_{ab} 
\bigg) \ ,
\ee 
with parameters
$\theta^{ab}$ of the form
$\theta^{ab} = \alpha m_{(\sigma)}^{ab}$
for some permutation $\sigma$.

We now argue that, for any $g\in SO(6)$,
the symmetry operator that realizes the transformation $g$ can be constructed by taking the fusion product
of a number of  $U(\theta = \alpha m_{(\sigma)})$ operators, for suitable $\alpha$, $\sigma$.
We also use the result, from Sec.~\ref{sec:fusion-section} and Appendix \ref{App:NonComm}, that the fusion of
$U(\theta)$ and $U(\theta')$
yields $U(\theta'')$ with
$\exp \theta \exp \theta' = \exp \theta''$.
It follows that the claim is mapped to a group theory statement, namely, 
that any $g\in SO(6)$ can be obtained as a product of factors of the form
$\exp(\alpha m_{(\sigma)})$,
for suitable $\alpha$, $\sigma$. To argue for this, we proceed in two steps.

First, we show how to realize an arbitrary rotation in the $(a,b)$-plane, for any 
$a$, $b$. For ease of exposition, we consider the $(1,2)$-plane,  the general case is analogous.
If we use the permutations
$\sigma = (123456)$ and $\sigma = (124365)$ with parameter $\alpha$, we realize the $SO(6)$ matrices
\begin{align}
    \exp(\alpha m_{(123456)})
&= \text{diag}\left(R_{12}(\alpha), R_{34}(\alpha),R_{56}(\alpha)\right), \\
\exp(\alpha m_{(124365)})
&= \text{diag}\left(R_{12}(\alpha), R_{34}(-\alpha),R_{56}(-\alpha)\right) \nonumber
\end{align} where $R_{ab} (\alpha)$ is the $SO(2)$ rotation matrix on the $(a,b)$-plane with angle $\alpha$.  The desired $(1,2)$-element of $SO(6)$ can be obtained by multiplying the two elements above.

Finally, having realized arbitrary rotations in the $(a,b)$-plane for any $a$, $b$, we can realize any element in $SO(6)$ by appealing to a generalization of $SO(3)$  Euler angles for $SO(n)$ \cite{genEuler}. 
 
\section{Noncommutative operator fusion}
\label{App:NonComm}
The operators in described in this paper were first constructed with a specific choice of maximal torus $T \cong U(1)^3$. It is true that such a choice exists for every element in $SO(6)$. However, to exhibit the full non-Abelian structure of the $SO(6)$ symmetry, one needs to relate the D5-KK bound states constructed with respect to different choices of maximal tori. In other words, there is a change of basis involved when we try to compose the action of two symmetry operators \eqref{boundstateEq}.

For concreteness, let us fix a reference maximal torus $T_0$. Any given element $g \in SO(6)$ is always related to some element $t \in T_0$ by conjugation, i.e.~$g = hth^{-1}$ for some $h \in SO(6)$, so a symmetry operator $U_g$ labeled by $g$ can be expressed similarly to \eqref{boundstateEq} as
\begin{equation}
    U_g = \exp\bigg(\frac{i}{2g^2} \, \alpha (hmh^{-1})^{ab} \int_{M^3} \ast F_{ab}\bigg) \, .\label{eq:generic_bound_state}
\end{equation}
Geometrically, this ``rotates'' the internal 2D subspace,
\begin{equation}
    \Sigma^2 \to \tfrac{1}{2} (hmh^{-1})^{ab} \Sigma^2_{ab} = \tfrac{1}{2} m^{ab} (h^{-1}  \Sigma^2 h)_{ab} \, .
\end{equation} 
The difference between the two expressions can be understood as active vs passive rotations. Importantly, the coefficients $\widetilde{m}^{ab} = (hmh^{-1})^{ab}$ can now be nonzero for generic $(a,b)$ unlike before, so we have effectively moved out of the maximal torus $T_0$.

The parallel fusion $U_g U_{g'} = U_{gg'}$ of two D5-KK bound states labeled by generic $g,g' \in SO(6)$ is noncommutative.
Let us see how this manifests at the level of the operators. 
Using $g = \exp(\alpha \widetilde{m})$ and $g' = \exp(\alpha' \widetilde{m}')$, the Baker-Campbell-Hausdorff formula gives 
\begin{widetext}
\begin{equation}
\begin{aligned}
    U_g U_{g'} & = \exp\bigg((\alpha \widetilde{m} + \alpha' \widetilde{m}')^{ab} X_{ab} + \frac{1}{2} \, \alpha \widetilde{m}^{ab} \alpha' (\widetilde{m}')^{cd} [X_{ab},X_{cd}] + \dots \bigg)\\
    & = \exp \bigg(\Big(\alpha \widetilde{m} + \alpha' \widetilde{m}' + \frac{1}{2} \, [\alpha \widetilde{m},\alpha' \widetilde{m}'] + \dots\Big)^{ab} X_{ab}\bigg) \ ,\label{eq:generic_fusion}
\end{aligned}
\end{equation}
\end{widetext}
where we have introduced the shorthand notation $X_{ab} =\frac{i}{2g^2} \int_{M^3} \ast F_{ab}$ for compactness. In the second step above, we noted that $X_{ab}$ satisfies the $\mathfrak{so}(6)$ algebra $[X_{ab},X_{cd}] = \tfrac 12f_{ab,cd}{}^{ef} X_{ef}$,
$f_{ab,cd}{}^{ef} = 8 \delta_{[a}{}^{[e} \delta_{b][c} \delta_{d]}{}^{f]}$.
The commutator terms are trivial if and only if $g$ and $g'$ commute.

Nonetheless, it is illuminating to consider a special case where $h^{-1} h'$ is in the centralizer of $(h')^{-1} g' h'$, then
\begin{equation}
    U_g U_{g'} = \exp\bigg(\frac{i(h(\alpha m + \alpha' m')h^{-1})^{ab} }{2g^2}  \int_{M^3} \ast F_{ab}\bigg) \, .\label{eq:simplified_fusion}
\end{equation}
Up to an overall conjugation by $h$, we see that the expression above recovers the discussion in Sec.~\ref{sec:fusion-section} regarding the fusion of hanging branes. By the same token, we also obtain a similar fusion rule with $h \to h'$ if $h^{-1} h'$ is in the centralizer of $h^{-1} g h$.

\bibliography{refs}

\end{document}